\documentstyle[12pt]{article}
 \newcommand{\r}{\rightarrow}
 \newcommand{\be}{\begin{equation}}
 \newcommand{\ee}{\end{equation}}
 \newcommand{\eel}[1]{\label{#1}\end{equation}}
 \newcommand{\bea}{\begin{eqnarray}}
 \newcommand{\eea}{\end{eqnarray}}
 \newcommand{\eeal}[1]{\label{#1}\end{eqnarray}}
 \newcommand{\baq}{\begin{equation}\begin{array}{rcl}}
 \newcommand{\eaq}{\end{array}\end{equation}}
  \newcommand{\eaql}[1]{\end{array}\label{#1}\end{equation}}
  \newcommand{\beac}{\begin{equation}\begin{array}{rcl}}
  \newcommand{\eeacn}[1]{\end{array}\label{#1}\end{equation}}
  \newcommand{\ba}{\begin{array}}
  \newcommand{\ea}{\end{array}}
  \newcommand{\non}{\nonumber \\}

  \renewcommand{\d}{\delta}

  \newcommand{\gym}{g_{YM}}

  \newcommand{\al}{{\alpha^{'}}}
  \newcommand{\beq}{\begin{eqnarray}}
  \newcommand{\eeq}{\end{eqnarray}}

  \newcommand{\w}{Schwarzschild $\:$}

  \newcommand{\bl}{\hspace{-.65cm}}
\begin{document}
  \begin{titlepage}
  
  \begin{flushright}
  hep-th/9809063\\UCSBTH-98-6
  \end{flushright}
  \vspace{2cm}
  
  \begin{center}
  \LARGE{D6+D0  and Five Dimensional\\ Spinning Black Hole}
  
  \vspace{10mm}
  
  \normalsize{ N. Itzhaki }\\
  
  \vspace{.7cm}
  \normalsize{
  Department of Physics\\
  University of California, Santa Barbara, CA 93106}\\
  
  \normalsize{\it sunny@solkar.physics.ucsb.edu}
  \end{center}
  \vskip 0.61 cm
  
  \begin{abstract}
  We study the system of D6+D0 branes at sub-stringy scale.  We show that
  the proper description of the system, for large background field associated 
  with the D0-branes, is via spinning chargeless black holes in five
 dimensions.
  The repulsive force between the D6-branes and the D0-branes is understood
  through the centrifugal barrier. We discuss the implication on the
  stability of the D6+D0 solution.
  
  \end{abstract} \end{titlepage} \baselineskip 18pt

  As it is well know there is a  repulsive force between  D6-branes and 
  D0-branes both at short and at large distances \cite{pol}.
  On the other hand a solution corresponding to a 
  background of D0-branes on D6-branes exist \cite{tay}.
  The background is stable classically, at least to quadratic order.
  Since super Yang-Mills (SYM) theory in seven dimensions is a 
  non-renormalizable theory one cannot test the stability of the background
  at the quantum level.
  The situation would have been different if there had been an underline
  theory 
  which flows at the IR to SYM.
  However, such a theory, which does not involve gravity,
 does not exist \cite{sen,sei,imsy,pp}.
  In this short note we take advantage of
  the recent progress in the 
  understanding of the relation between the
  near horizon geometry of a given
  branes configuration and the field theory living
  on the branes \cite{mal}
  and study the near horizon geometry of D6-branes with a constant field
  associated with the D0-brane.
  We consider the ``decoupling'' limit while keeping the super-Yang-Mills
  coupling constant and the field strength, 
  associated with the D0-brane background, fixed.
  We find that the near horizon geometry is that of a spinning black hole
  in five
  dimensions.
  When the field strength is large (compared to $\gym^{-4/3}$)
  the size of the black hole horizon 
  is large (compared to the Planck scale)
  and hence the supergravity solution can be trusted in the analysis of  
  the stability of the D0-branes background. 
  
  Before we consider the D6+D0 system let us review the decoupling limit
  of D6-branes.
  The  ``decoupling'' limit is defined as follows \cite{juan}
  \be\label{limD6} 
  U=\frac{r}{\al}=\mbox{fixed},\;\;~~~~
  \gym^2=(2\pi)^4 g_s\al^{3/2}=\mbox{fixed},\;\;\;~~~~
  \al\r 0.
  \ee
  In this limit we keep the field theory energies and coupling constant fixed
  while taking $\al$ to zero.
  This suggest that the D6-branes decouple from the bulk.
  However, as was noticed in \cite{sen, sei} in this limit $R_{11}\propto
1/\al
  \r\infty$, which means that the right description of the system is in M-theory
  as an ALE space with $A_{N-1}$ singularity (where $N$ is the number of
  D6-branes).
  Note that in this limit the Planck length is finite \cite{sen,sei}
  \be
  l_p=(2\pi)^{-4/3}\gym^{2/3}.
  \ee
  To be more precise we can start with the type IIA solution
  associated with D6-branes \cite{hs} and take the limit (\ref{limD6})
  to obtain \cite{imsy}
  \beq\label{solD6}
  && ds^2=\al\left( \frac{(2\pi)^2 }{\gym}\sqrt{\frac{2 U}{ N}}dx^2_{||}+
  {\gym \over (2\pi)^2 }
  \sqrt{\frac{N}{2 U}}dU^2+{\gym \over (2\pi)^2 \sqrt{2}}
  \sqrt{N}U^{3/2}d\Omega^2\right), \non
  && e^{\phi}= { g^2_{YM} \over 2 \pi} 
  \left(2  \frac{U}{\gym^2N}\right) ^{3/4}.
  \eeq
  The solution  can be trusted in the region
  $\frac{1}{\gym^{2/3}N^{1/3}}\ll U\ll \frac{N}{\gym^{2/3}}$ where
  both the curvature (in string units) and the effective string 
  coupling are small \cite{imsy}.
  
  For large $U$ the effective string coupling becomes large and we need to 
  uplift the solution to eleven dimensions to obtain,
  \be\label{kkmonshor}
  ds^2=dx^2_{||}+\frac{l_p^3N}{2 U}dU^2+{ l_p^3NU \over 2} (d
  \tilde{\theta}^2+
  \sin^2 \theta d\varphi^2)
  +\frac{2 U l_p^3 }{N}[d\phi+ { N \over 2 } (\cos\tilde{\theta} -1)
  d\varphi ]^2
  \ee
  where $\phi \equiv x_{11}/R_{11}$ has period $\phi \sim \phi + 2 \pi$. 
  Defining the new variables 
  \beq\label{cc}
  y^2 = 2 N l_p^3 U ,~~~~~~   \theta = \tilde\theta /2, ~~~~~~
  ~ \phi_1 = \varphi + \phi/N, ~~~~~ ~ \phi_2 = \phi/N
  \eeq 
  we get the metric
  \be\label{An}
  ds^2=dx^2_{||}+ dy^2 + y^2 (d \theta^2 +
  \sin^2\theta d\phi_1^2 +
  \cos^2\theta d \phi_2^2 ),
  \ee
  where $0\leq \theta \leq \pi/2$ and $0 \leq  \phi_1 ,
  \phi_2 \leq 2\pi$ with the identification $(\phi_1, 
  \phi_2) \sim (\phi_1, \phi_2) + ( 2\pi/N, 2\pi/N)$.
  This identification leads  to an 
  ALE space with an $A_{N-1}$ singularity.

  Moreover, starting with near-extremal D6-branes with finite energy density
  above extremality we end up with a five dimensional
  \w black hole sitting at the $A_{N-1}$ singularity times $R^6$
  \cite{imsy}
  \be
  ds^2=- ( 1 - {y_0^2 \over y^2 }) dt^2 + { dy^2 \over 
  ( 1 - {y_0^2 \over y^2 }) }
  + y^2  d \Omega_3^2 + dx_i^2
  \ee
  where $i=1,....,6$.
  
  Now we wish to add D0-branes or in the field theory language we wish to 
  find the supergravity solution associated with the D0-branes background
  of \cite{tay}.
  Namely, we keep the energy of the D0-branes background
  fixed while taking the limit (\ref{limD6}).
  In M-theory D0-branes are described by 
  gravitational waves along the $x_{10}-x_0$ direction.
  Thus they carry energy which contributes to the total mass
  of the black hole solution.
  Since they also carry momentum along $x_{10}$ and since at the near horizon 
  geometry of the D6-branes $x_{10}$ is related to $\phi$ via
  $\phi=x_{10}/R_{10}$, the D0-branes 
  will contribute also angular momentum to the black hole.
  This implies that the near horizon geometry of D6+D0 system is that of a 
  spinning  black hole.
  In fact, since from the 11D point of view
  D6-branes and D0-branes excite  only the 
  metric fields the solution is that of a chargeless black hole.
  Chargeless black holes in five dimensions are described by three parameters:
  the mass, the angular momentum in the,  $x_7, x_8$ plane and the angular
  momentum in the $x_9, x_{10}$ plan \cite{mp}.
   From   eq.(\ref{cc}) it is clear that in our case 
  \be\label{jj} 
  J_{7,8}=J_{9,10}\equiv J.
  \ee
  The solution is therefore \cite{mp},
  \beq\label{lim}
  && 
  ds^2=-dt^2+\frac{(y^2+a^2)^3}{(y^2+2a^2)(y^2+a^2)^2-\mu y^2}dy^2
  \non &&
  +\frac{\mu y^2(y^2+2a^2)}{(y^2+a^2)^3}(dt+a \sin^2\theta d\phi_1
  +\cos^2\theta d\phi_2)^2
  \\
  &&
  +(y^2+2a^2)(d\theta^2+\sin^2\theta d\phi_1^2+\cos^2\theta d\phi_2^2).
  \nonumber
  \eeq
  One way to check that this is indeed the near horizon geometry of 
  D6+D0 branes is to use eq.(\ref{cc}) backwards 
  and then reduce the solution along the
  $\phi$ direction  to ten dimensions.
  Schematically (for more details see the appendix)  one gets
  \be\label{gg}
  ds^2=A(U)(d\phi+\frac{N}{2}(\cos\theta -1)d\varphi+B(U)dt)^2+\tilde{ds}^2,
  \ee
  where $\tilde{ds}^2$ does not depend on $\phi$ and $d\phi$.
  We see, therefore, that an electric charge, associated with the D0-branes,
  appears and that
  the magnetic charge, associated with the D6-branes, is the same as in the 
  solution with no D0-branes, as it should. 
  Note that (\ref{jj}) is crucial to obtain 11D solution of the form (\ref{gg}).

  The mass and angular momentum associated with this solution can be read from
  the asymptotic behavior of the solution and yields \cite{mp},
  \beq\label{mj}
  && M=\frac{3 \pi \mu}{8 G},\non
  && J=\frac{2}{3}Ma,
  \eeq
  where $G$ is the five dimensional Newton constant.
  When  the only source of energy is the D0-branes background (no 
  additional thermal energy) the solution is extremal ($4a^2=\mu$)
  and, hence, it  does not Hawking radiates.
  This means that the D0-branes background  is stable at the
  quantum level as well.
  To learn about the nature of this stability one can add some thermal 
  noise on-top of the D0-branes background.
  By doing so one gets  a non-extremal spinning five dimensional
  black hole ($4a^2<\mu$).
  Such a black hole will Hawking radiates the angular momentum  
  before it radiates the energy above extremality. 
   From the field theory point of view  this means that once we add some
  amount of energy on top of the D0-branes  background the background
  becomes non-stable.
  This is, of course, in agreement with the fact that there is a repulsive 
  force between the D0-branes and the D6-branes.
  
  As we have seen, at the near horizon geometry of D6-branes the D0-branes
  contribute to the
  angular momentum.
  The repulsive force which they fill is, therefore,
  simply the centrifugal barrier,
  which in five dimensions has the form
  \be
  V\propto \frac{L^2}{y^2}\propto \frac{L^2}{U}.
  \ee 
  It is interesting to note that the same behavior was found in \cite{gil}.
  Note, however, that the regions of validity of the computations are
  different.
  In \cite{gil} there are two kinds of computations. The first uses the
  D-branes technique  which is valid
  at the sub-stringy region. The second is based on the supergravity solution
  at large distances compared to the string scale
 (where the $1$ in the harmonic function is kept).
  Our approach, in the spirit of \cite{mal},
  is valid at the sub-stringy region but it
  relies on supergravity.
  To trust the supergravity solution at the sub-stringy region we need the
  background field to be large, 
  while to trust the sub-stringy computation of \cite{gil} one needs the
  background field to be small \cite{imsy}.
  The fact that the potential is insensitive to the 
  interpolation between the small and the large 
  background field is in agreement with  the result of 
  branes probing for this configuration \cite{pie,bisy,bra,dm}.
  
  \vspace{7mm}

\bl I would like to thank S. Yankielowicz for helpful comments.

\vspace{7mm}

  \bl { \bf Appendix}
 
  In this appendix we discuss in more details the relation between $n_0$,
  the number of D0-branes,  and the five dimensional
   and the angular momentum of the black hole.
  
  The type IIA solution of D6+D0 was presented in \cite{bisy}\footnote{The
  self dual solution can be found also in \cite{ort,she}.}.
  This solution is a simple generalization of the four dimensional solution 
  of \cite{gw}.
  In the large $U$ region, which corresponds to the large $y$ region
  (where the 
  mass and angular momentum of the black hole
are defined), the gauge field part of the 
  solution in the limit (\ref{limD6}) is 
  \be\label{opo}
  A_{\mu}dx^{\mu}=\frac{\sqrt{3}QN}{8\al^2 U^2}dt+\frac{N}{2}(1-\cos
  \tilde\theta ) \d\phi,
  \ee
  where,
  \be\label{Q}
  Q=\frac{g_s n_0 (2\pi)^6 \al^{7/2}}{2V_6}.
  \ee
  Note that in this limit $A_0$ is proportional to $ 1/U^2$ and not to $1/U$.
  The reason is that we are in the large $U$ region but not at the large
  $r$ region.
  The full solution \cite{bisy} in the large $r$ region yields $A_0\propto
  1/r$.\footnote{Similar phenomena occurs also in the localized solution of 
the D2-branes within D6-branes \cite{ity}.
}

 Since  $F_0$ is held fixed  while taking the limit (\ref{limD6})
  and since the number of D0-branes on the D6-branes is given by \cite{tay}
  \be
  n_0=\frac{1}{6(2\pi)^6}\int d^6x \mbox{Tr}F\wedge F\wedge F
  \ee
   $n_0$ is also fixed  in this limit.
This implies that $Q\propto \al^2$
  (where we have used eqs.(\ref{Q}, \ref{limD6}) and that $A_0$ is fixed , as expected.
  
Starting from the spinning black hole solution and reducing it to 10D
  one finds that
  \be\label{pop}
  A_0=\frac{\mu a}{y^4}.
  \ee
 Comparing to (\ref{opo}) one finds  the right $y$ dependence
 (since $y^2\propto U$, eq.(\ref{cc})). 
Moreover, eqs.(\ref{mj}, \ref{opo}, \ref{pop}) also verify that
  \be
  J\propto n_0.
  \ee


\begin{thebibliography}{99}
  
  
  \bibitem{pol}J. Polchinski, {\it TASI Lectures on D-branes}, hep-th/9611050.
  
  \bibitem{tay} W. Taylor, {\it Adhering Zero-Branes To Six-Branes and 
  Eight-Branes}, Nucl. Phys. B508 (1997) 122, hep-th/9705116.
  
  
  
  
  \bibitem{sen} A. Sen, {\it D0 Branes on $T^n$ and Matrix Theory}, 
  Adv. Theor. Math. Phys.2 (1998) 51, hep-th/9709220.
  
  \bibitem{sei} N. Seiberg, {\it Why Is The Matrix Model Correct},
  Phys. Rev. Lett. 79 (1997) 3577, hep-th/9710009.
  
  \bibitem{imsy} N. Itzhaki, J.M. Maldacena, J. Sonnenschein and S. Yankielowicz,
  {\it Supergravity and The Large N Limit of Theories With Sixteen
  Supercharges},
  Phys. Rev. D58 (1998) 046004, hep-th/ 9802042.
  
  \bibitem{pp} A.P. Peet and J. Polchinski,
  {\it UV/IR Relations in AdS Dynamics}, hep-th/9809022.
  
  \bibitem{mal} J.M. Maldacena, {\it The Large N Limit of Superconformal
  Field Theories and Supergravity}, hep-th/9711200.
  
  \bibitem{juan} J.M. Maldacena, {Branes probing black holes},
  Nucl. Phys. Proc. Suppl. 68 (1998) 17, hep-th/9709099.
  
  \bibitem{hs} G. Horowitz and A. Strominger, {\it Black Strings and p-branes},
  Nucl. Phys B360 (1991) 197.
  
  \bibitem{mp} R.C. Myers and M.J. Perry, {\it Black Holes 
  in Higher Dimensional Space-Time}, Ann. Phys. 172 (1986) 304.
  
  \bibitem{gil} G. Lifschytz, {\it Comparing D-branes To
  Black-branes},  Phys. Lett. B388 (1996) 720, hep-th/9604156. 
  
  \bibitem{pie} J.M. Pierre, {\it Comparing D-branes and Black Holes
  with 0- and 6-brane},  Phys. Rev. D56 (1997) 6710, hep-th/9707102.
  
  \bibitem{bisy} A. Brandhuber, N. Itzhaki, J. Sonnenschein and S.
 Yankielowicz,
  {\it More on Probing Branes with Branes}, Phys. Lett. B423 (1998) 238, 
  hep-th/9711010. 
  
\bibitem{bra} J. Branco, {\it  Probing a D6 + D0 state with D6-branes:
  SYM - Supergravity correspondence at subleading level},
  hep-th/9806186. 
  
  
  \bibitem{dm} A. Dhar and G. Mandal, {\it Probing 4-Dimensional 
  Nonsupersymmetric Black Holes Carrying D0- and D6-brane charges},
  hep-th/9803004. 
  
  
\bibitem{ort} R. Khuri and T. Ortin, {\it A Nonsupersymmetric
  Dyonic Extreme Reissere-Nordstrom Black Hole}, Phys. Lett. B373 (1996) 56,
  hep-th/9512178.
  
  \bibitem{she} H. J. Sheinblatt, {\it Statistical Entropy of an Extremal
  Black Hole with 0- and 6-Brane Charge},
  Phys. Rev. D57 (1998) 2421,hep-th/9705054.
  
  
  \bibitem{gw} G.W. Gibbons and D.L. Wiltshire, {\it Black Holes
  in Kaluza-Klein Theory},  Ann. Phys. 167 (1986) 201.,
  ERRATUM-ibid.176 (1987) 393.
  
\bibitem{ity} N. Itzhaki, A.A. Tseytlin and S. Yankielowicz,
    {\it Supergravity Solutions for Branes
Localized Within Branes}, Phys. Lett. B432 (1998) 298, hep-th/9803103.
  
  
  \end{thebibliography}
  \end{document}